\documentclass[aps,prl,reprint,superscriptaddress]{revtex4-2}
\usepackage[usenames,dvipsnames]{color}
\newcommand{\red}[1]{\textcolor{black}{#1}}

\usepackage{graphicx}
\usepackage{xcolor}
\usepackage[version=4]{mhchem}
\begin{document}
%Title of paper
\title{Radiative stabilization of the indenyl cation: Recurrent fluorescence in a closed-shell polycyclic aromatic hydrocarbon}

\author{James N. Bull}
\email{james.bull@uea.ac.uk}
\affiliation{School of Chemistry, University of East Anglia, Norwich NR4 7TJ, United Kingdom}
\affiliation{Centre for Photonics and Quantum Science, University of East Anglia, Norwich NR4 7TJ, United Kingdom}
\author{Arun Subramani}
\affiliation{Department of Physics, Stockholm University, SE-10691 Stockholm, Sweden}
\author{Chang Liu}
\affiliation{School of Chemistry, University of Melbourne, Parkville, VIC 3010, Australia}
\author{Samuel J. P. Marlton}
\affiliation{School of Chemistry, University of Melbourne, Parkville, VIC 3010, Australia}
\author{Eleanor K. Ashworth}
\affiliation{School of Chemistry, University of East Anglia, Norwich NR4 7TJ, United Kingdom}
\author{Henrik Cederquist}
\affiliation{Department of Physics, Stockholm University, SE-10691 Stockholm, Sweden}
\author{Henning Zettergren}
\affiliation{Department of Physics, Stockholm University, SE-10691 Stockholm, Sweden}
\author{Mark H. Stockett}
\affiliation{Department of Physics, Stockholm University, SE-10691 Stockholm, Sweden}
\begin{abstract}
Several small polycyclic aromatic hydrocarbons (PAHs) with closed-shell electronic structure have been identified in the cold, dark environment Taurus Molecular Cloud-1. We measure efficient radiative cooling through the combination of recurrent fluorescence (RF) and IR emission in the closed-shell indenyl cation (C$_{9}$H$_{7}^{+}$), finding good agreement with a master equation model including molecular dynamics trajectories to describe internal-energy dependent properties for RF. We find that C$_{9}$H$_{7}^{+}$ formed with up to $E_{c}=5.85$\,eV vibrational energy, which is $\approx$2\,eV above the dissociation threshold, radiatively cool rather than dissociate. The efficient radiative stabilization dynamics are likely common to other closed-shell PAHs present in space, contributing to their abundance.
\end{abstract}

\maketitle

%\section{Introduction}
The discovery of small polycyclic aromatic hydrocarbon (PAH) molecules in interstellar clouds through radioastronomy \cite{McGuire2021,Burkhardt2021,Sita2022,Cernicharo2021,Cernicharo2024,Wenzel2024,Wenzel2024a} and observations of infrared (IR) emission bands of PAHs with unprecedented sensitivity by the James Webb Space Telescope (JWST) \cite{GarciaBernete2022,Spilker2023,Witstok2023}, is opening up a new era for laboratory investigations of PAH unimolecular dynamics. The general consensus is that PAHs are widespread in space \cite{Tielens2008}, with data from the Spitzer Space Telescope and JWST supporting that up to 20\% of galactic carbon is present as PAHs \cite{Peeters2011,Li2020,GarciaBernete2022,Chabot2019,Witstok2023}. To date, fewer than ten specific PAH-based molecules have been discovered in space. In all cases, identifications were made in the Taurus Molecular Cloud-1 (TMC-1) through comparison of observed rotational lines with spectra recorded in the laboratory \cite{Burkhardt2021,Sita2022,Cernicharo2021,Cernicharo2024}. However, the observed abundances of these PAHs (1-cyanonaphthalene, 2-cyanonaphthalene, indene, 2-cyanoindene, 1/5-cyanoacenaphthylene and isomers of cyanopyrene) are orders of magnitude higher than astrophysical models predict, implying that the efficiencies of destruction channels are overestimated, and/or that the efficiencies of molecular formation, and  stabilization processes are underestimated.

In the cold molecular cloud where these specific PAHs were discovered, molecules are mostly shielded from the harsh interstellar radiation field \cite{Fuente2019}. The main destruction mechanisms included in astrochemical models are thermal ($\approx$10\,K) collisions with atomic and molecular cations \cite{Wakelam2008,Wakelam2015,Burkhardt2021,McGuire2021,Doddipatla2021,Yang2023}. These models do not consider if the resulting PAH cations may be reneutralized through electron-ion recombination or by mutual neutralization in collisions with anions, in addition to not including efficient (barrierless) ion-molecule bottom-up formation mechanisms, with recombination leading to the observed neutral PAHs \cite{McGuire2021,Burkhardt2021}. All of these reactions may generate neutral PAHs with internal energies similar to or in excess of their respective dissociation thresholds. It is therefore necessary to understand how energized closed-shell PAHs may stabilize through radiative cooling.

In a step toward investigating the radiative cooling processes contributing to PAH lifecycles, laboratory studies on isolated open-shell (radical cation) PAHs using electrostatic ion-beam storage devices \cite{Bernard2017,Martin2019,Saito2020,Stockett2023,Bernard2023,Lee2023} -- including under `molecular cloud in a box conditions' (\textit{P}$\approx$10$^{-14}$\,mbar, \textit{T}$\approx$13\,K) at the DESIREE infrastructure \cite{Thomas2011,Schmidt2013} -- have shown that recurrent fluorescence (RF), or Poincar\'{e} fluorescence \cite{Leger1988,Ebara2016a} plays a crucial role. Similar experiments have shown that RF dominates the cooling of odd-electron carbon chain anions and fullerene ions \cite{Ebara2016a,Ito2014,Andersen1996,Tomita2001,Chen2019,Chandrasekaran2014}. However, while ionized PAHs are expected to exist in bright or star-forming regions of space \cite{Montillaud2013}, it is their neutral (closed-shell) forms that have been observed in dark interstellar clouds \cite{Burkhardt2021,Sita2022,Cernicharo2021}. Closed-shell PAHs typically have different electronic structure than their open-shell cations \cite{Platt1949,Yang2016}. In particular, they typically have higher transition energies (by 1\,eV or more) to their first few electroically excited states; therefore, from results on open-shell PAH cations, one can not infer directly that RF-processes will be similarly active for closed-shell PAHs. From the perspective of laboratory experiments, it is not possible to store neutral PAHs, or investigate radiative cooling dynamics, over timescales relevant for determining their decay properties (microseconds to minutes). Thus, it is necessary to find other ways to investigate the balance between radiative stabilization and decay (fragmentation) of internally hot, closed-shell PAHs.

\begin{figure*}[!t]
\begin{center}
\includegraphics[scale=0.75]{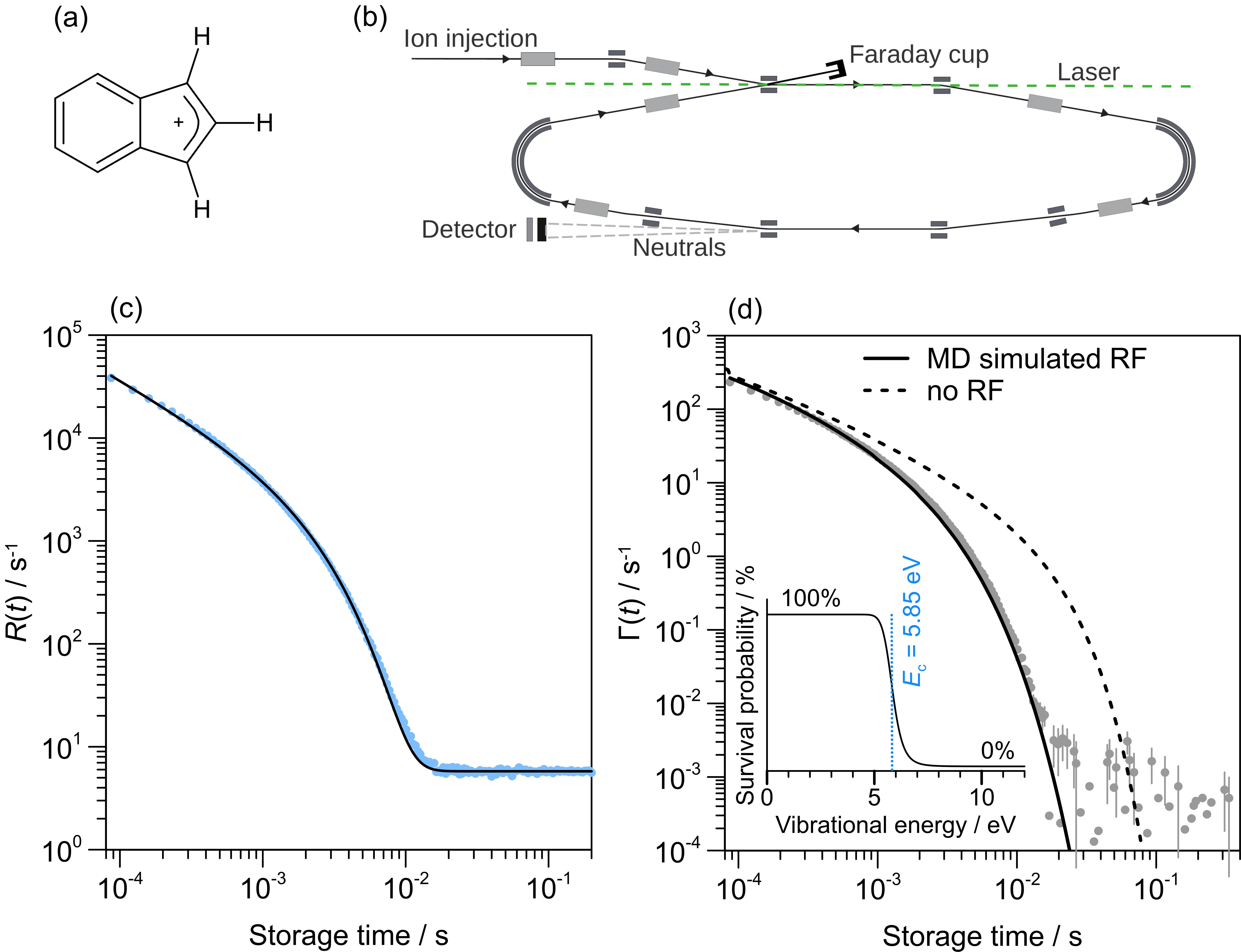}
\end{center}
\caption{Radiative cooling of energized [In-H]$^{+}$: (a) Molecular structure of [In-H]$^{+}$ (C$_{9}$H$_{7}^{+}$). (b) Illustration of one of the ion-beam storage rings at DESIREE, highlighting the important components for this work -- see Supplementary Material \cite{SM} for details of experimental methods. (c) Measured spontaneous neutral dissociation product curve (log-log axes) -- uncertainty bars are small compared to the size of the symbols. The black trace is a fit of the data to Eq.\,\ref{eqn:1}. (d) Absolute per-particle dissociation rate with ion storage time (grey points), $\Gamma(t)$, and the corresponding master equation model results with (solid black trace) and without (dashed trace) molecular-dynamics-simulated RF contributions. Experimental error bars are smaller than the symbols in most cases. \red{For $R(t)$ at} long storage times (i.e. $>$2$\times$10$^{-1}$\,s), a weak dissociation signal \red{($\approx$5.8\,s$^{-1}$)} is observed from collisions with residual gas (i.e. associated with the beam storage lifetime)\red{, and was subtracted when determining $\Gamma(t)$}. The inset in (d) shows survival probability of the ions due to RF, determined as $\frac{k_{RF}}{k_{RF}+k_{d}}\times 100\%$.}
\label{fig:1}
\end{figure*}

Here, we present experimental and theoretical results on highly efficient radiative cooling of the indenyl cation ([In-H]$^{+}$, C$_{9}$H$_{7}^{+}$, FIG.\,\ref{fig:1}a). This ion has a closed-shell electronic structure and transition probabilities similar to neutral PAHs with 20 or more carbon atoms (Supplementary Material \cite{SM}). In TMC-1, [In-H]$^{+}$ may be formed in ionizing collisions between indene and atomic cations, which is assumed to be the major destruction pathway for PAHs in there \cite{McGuire2021}. Outside of cold molecular clouds, [In-H]$^{+}$ may form with near unit quantum yield following the dissociative ionization of indene (C$_{9}$H$_{8}$) by $h\nu=13.6$\,eV (H-ionization) radiation \cite{West2014}. Importantly, to date, indene is the only pure PAH that has been observed in TMC-1 or in space. Therefore, information on the stabilities and reactivities of its precursors as well as of its fragments\red{, such as C$_{9}$H$_{7}^{+}$}, is crucial for estimates of abundances \cite{Burkhardt2021}. 

The present experiments were performed at the electrostatic ion-beam storage ring DESIREE \cite{Thomas2011,Schmidt2013} as shown in FIG.\,\ref{fig:1}b. There, we monitored the generation of neutral particles following thermal- or photo-activation of the stored ions in a 8.6\,m circumference storage ring (Supplementary Material \cite{SM}). Cryogenic cooling of the storage ring to $T\approx 13$\,K gives a residual gas density of $\approx$10$^{4}$ particles/cm$^{3}$ (mostly H$_{2}$) \cite{Schmidt2013}, allowing ions to be stored in a nearly collision-free environment for hours \cite{Baeckstroem2015}. For \red{ensembles of internally-excited C$_{9}$H$_{7}^{+}$, we measured the $R(t)$ of neutral particles (dissociation products) leaving the ensemble and ring as a function of the time $t$ after injection. At longer times, $R(t)$ is quenched due to radiative cooling of the stored ions.} Schematic illustrations of $R(t)$ data under different cooling scenarios are given in the Supplementary Material \cite{SM}. The measured $R(t)$-distribution is shown in FIG.\,\ref{fig:1}c, with the dissociation signal dominated by generation of C$_{2}$H$_{2}$ (giving \emph{m/z}=89 as the cyclopentadienyl acetylene cation and benzocyclopropyl cation isomers in a 2:1 ratio as confirmed through ion mobility spectrometry \red{-- see Supplementary Material \cite{SM}}), when vibrational energies are within a few electronvolts of the dissociation threshold. Loss of C$_{4}$H$_{2}$ and C$_{4}$H$_{4}$ becomes important at higher vibrational energies beyond the range considered in this work (TABLE\,S2). $R(t)$ is well-described by the expression

\begin{equation}
R(t) = r_{0}t^{-\alpha}e^{-k_{c}t}
\label{eqn:1}
\end{equation}

\noindent where $r_{0}$ is the initial signal amplitude and $k_{c}$ is a characteristic rate. Parameter $k_{c}^{-1}$ is the characteristic time at which dissociation is quenched by radiative emission. Eq.\,1 assumes that the count rate at short time follows a power law, $R(t) \propto t^{-\alpha}$ \cite{Hansen2001}, because of the broad vibrational energy distribution, $g(E, t)$, and the rapid variation of the dissociation rate coefficient, $k_{d}(E)$, with the internal excitation energy, $E$ \cite{Andersen2001}. A fit of Eq.\,\ref{eqn:1} to our $R(t)$ data (the black curve in FIG.\,\ref{fig:1}c) returned $k_{c}$=508$\pm$5\,s$^{-1}$ and $\alpha$=0.787$\pm$0.005, where $k_{c}$ is larger than for radical cations of naphthalene (460$\pm$5\,s$^{-1}$) \cite{Lee2023}, 1-cyanonaphthalene (300$\pm$20\,s$^{-1}$) \cite{Stockett2023}, and the $k_{c}$ values for larger PAHs such as anthracene, tetracene, or perylene \cite{Martin2013,Bernard2023,Stockett2020b}. Importantly, \red{photons from RF processes have} been directly observed for naphthalene and anthracene cations \cite{Saito2020,Kusuda2024}, highlighting that RF is a necessary factor for the cooling dynamics in PAH radical cations with large $k_{c}$ values. Typically, $k_{c}$ for systems with only IR cooling are $\sim$100\,s$^{-1}$ \cite{Zhu2022a}. We conclude that radiative cooling is particularly efficient in [In-H]$^{+}$.

\begin{figure}[!t]
\begin{center}
\includegraphics[scale=0.8]{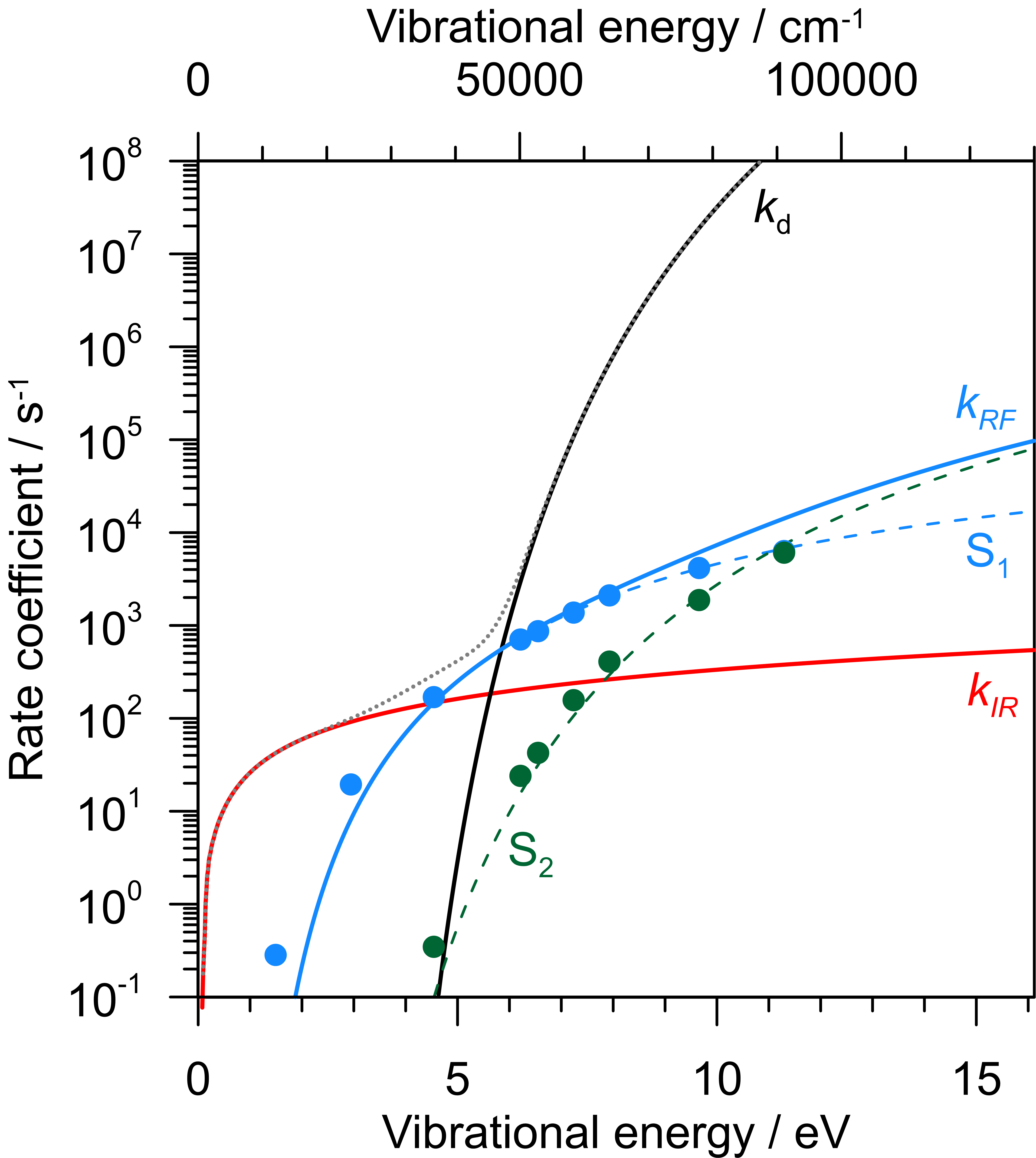}
\end{center}
\caption{\red{Calculated rate coefficients for dissociation ($k_{d}$), infrared emission ($k_{IR}$), and recurrent fluorescence ($k_{RF}$). Note that $k_{d}$ and $k_{RF}$ cross at $\approx$5.85\,eV. The $k_{RF}$ curve considers both S$_{1}\rightarrow$\,\,S$_{0}$ and S$_{2}\rightarrow$\,\,S$_{0}$ fluorescence transitions (dashed traces), although only the former is important for radiative cooling for the internal energies considered here. The solid points are time-averaged rate coefficients determined from AIMD trajectories at specific, fixed internal energies -- see Supplementary Material \cite{SM}. The RF curves (dashed lines) were determined from Eq.\,3, in the Supplementary Material \cite{SM}, and use oscillator strengths from a fit across the AIMD data. Thresholds are 1.20\,eV for RF and 3.74\,eV for dissociation.}}
\label{fig:2}
\end{figure}

While a large value of $k_{c}$ implies efficient radiative cooling, it does not yield any information on the detailed relaxation mechanisms. We use experimental parameters to convert the $R(t)$ data to absolute rates per molecule, $\Gamma(t)$, as shown in FIG.\,\ref{fig:1}d (Supplementary Material \cite{SM}). We model the rates, shown in \red{FIG.\,\ref{fig:2}}, for dissociation ($k_{d}$) using RRKM theory; RF emission ($k_{RF}$) via internal-energy-dependent geometries and associated transition energies and oscillator strengths obtained from \emph{ab initio} molecular dynamics (AIMD) trajectories; and IR emission ($k_{IR}$) using a simple harmonic cascade framework \cite{Stockett2023,Lee2023,Bull2019}. These three processes were combined in a coupled master equation to model $\Gamma(t)$ (black trace in FIG.\,\ref{fig:1}d). The model results are in good agreement with the experimental data. 

\red{A key outcome from the modeling is that} if RF is excluded, we arrive at the dashed curve in FIG.\,\ref{fig:1}d, which substantially overestimates the dissociation rate. \red{Moreover,} use of static oscillator strengths at the molecular equilibrium geometry (i.e. assuming \textit{T}$=$0\,K) for RF gave an observed dissociation curve close to the `no RF' case, showing that internal-energy-dependent (i.e. temperature dependent) effects leading to geometric distortions\red{, which were incorporated here using AIMD trajectories,} are critical for a realistic description of the RF process. \red{Indeed, while earlier studies on open-shell PAH cations have suggested that $k_{RF}$ may be significantly underestimated by static calculations of $k_{RF}$ and sought to include Herzberg-Teller coupling \cite{Herzberg1933} at the equilibrium geometry to more robustly describe oscillator strengths \cite{Stockett2023,Lee2023}, the use of AIMD trajectories to determine internal-energy-dependent transition energies and probabilities provides a decisive step in describing RF.} \red{The present AIMD trajectory strategy is} readily applied to other open- and closed-shell PAHs.

\red{Taking the $k_{RF}$ and $k_{d}$ rates from FIG.\,\ref{fig:2},} the modeled survival probability for [In-H]$^{+}$ as a function of vibrational energy is shown in the inset in FIG.\,\ref{fig:1}d, returning the critical vibrational energy $E_{c}$=5.85\,eV, above which the ions are more likely to be destroyed (fragmented) and below which they are more likely to be stabilized through radiative emission. Significantly, $E_{c}$ is $\approx$2\,eV higher than the lowest bond dissociation energy (BDE), which we calculated here to be 3.74\,eV (Supplementary Material \cite{SM}). This clearly demonstrates the importance of RF for \red{the stabilization of the} closed-shell PAH cation [In-H]$^{+}$.

We now consider how the radiative cooling dynamics observed for [In-H]$^{+}$ relate to neutral PAHs that may be present in dark molecular clouds, which, like [In-H]$^{+}$, have closed-shell electronic structures. Based on the close agreement found between theory and experiment for \red{the cooling of} [In-H]$^{+}$, and established trends in key molecular properties of PAHs with increasing size, predications on the rates for dissociation, IR emission, and RF \red{for other PAHs are} possible. First, dissociation rates within an RRKM framework decrease with increasing PAH size (i.e., heat capacity) because, while exceptions will exist, H-loss dissociation energies show little variation across PAHs containing 10--24 carbon atoms \cite{Simon2017,Reitsma2014}, with astrochemical models often assuming a value of 4.1--4.6\,eV \cite{Ling1995,Andrews2015}. Similarly, because IR emission rates are usually dominated by a few bright modes that are common across PAHs as a class, the total IR emission rate slowly decreases with increasing size and heat capacity \cite{Montillaud2013}. On the other hand, while the first electronic transition energies of closed-shell PAHs have some dependence on the chemical structure, the overall trend is that the corresponding excitation energies decrease with increasing PAH size \cite{Vijh2004}, which favors stabilization through RF and counteracts the effect of increasing heat capacity. For example, transition energies in acenes containing 10 to 20 carbon atoms, including non-linear, 2D, and structural isomers, systematically decrease from $\approx$3 to 1.5\,eV \cite{Parac2003,Halasinski2003,Malloci2011,Yang2016}, approaching that of [In-H]$^{+}$ at 1.2\,eV (S$_{1}$), but also have oscillator strengths that are larger than [In-H]$^{+}$ by at least an order of magnitude. This indicates that RF will tend to become more important with increasing PAH size, however, computational studies on a range of neutral PAHs using our (or similar) modeling framework are desirable to more precisely quantify the degree to which these trends scale.

This work has demonstrated efficient radiative cooling in a small, closed-shell PAH molecule, in which RF plays a dominant role in quenching dissociation for internal energies up to $E_{c}$=5.85\,eV. Our target system, C$_{9}$H$_{7}^{+}$, is of similar size to the smallest neutral, closed-shell PAHs observed in TMC-1 through radioastronomy, suggesting that other similarly sized, closed-shell PAHs also may have efficient cooling dynamics. The enhanced survivability of energized neutral PAHs with efficient RF cooling means that the abundance of small, closed-shell PAHs (i.e., the building blocks in bottom-up growth astrochemistry) inherited from earlier stages of molecular cloud formation may be higher than expected \cite{McGuire2021,Burkhardt2021,Sita2022,Cernicharo2021,Cernicharo2024,Wenzel2024,Wenzel2024a}. Ultimately, this work demonstrates that RF must be considered in astrophysical molecular abundance models in order to reliably predict contributions to aromatic infrared bands (AIBs) and other IR observations for closed-shell PAHs \cite{Tielens2008,Lacinbala2023}, and that the RF influence will become gradually more important for larger PAHs. The exceptionally good agreement found here for the measured dissociation rate between experiment and theory for a prototype closed-shell PAH means that our cooling model, combining internal-energy-dependent transition energies and oscillator strengths, for describing RF can be applied to predict the radiative cooling rates of other closed-shell PAH molecules.

\section{acknowledgements}
Funding was provided by the Swedish Foundation for International Cooperation in Research and Higher Education (STINT) Grant for Internationalisation programme (PT2017-7328), an EPSRC New Investigator Award (EP/W018691), the Knut and Alice Wallenberg Foundation  (2018.0028, Probing charge- and mass-transfer reactions on the atomic level), and the Swedish Research Council (contract no. 2019-04379, 2018-04092, 2020-03437, and 2023-03833). This work was performed at the Swedish National Infrastructure, DESIREE, supported by the Swedish Research Council contract no. 2017-00621 and 2021-00155. EKA thanks the University of East Anglia for doctoral studentship. Antony Hinchliffe, Manager of the Mass Spectrometry Platform at UEA, is thanked for technical assistance on recording ion mobility and CID data. Electronic structure calculations were, in part, carried out on the High Performance Computing Cluster supported by the Research and Specialist Computing Support service at the University of East Anglia. This article is based upon work from COST Action CA18212 -- Molecular Dynamics in the GAS phase (MD-GAS), supported by COST (European Cooperation in Science and Technology). The authors thank the DESIREE operators for their unwavering support.

\bibliography{indenyl}

\end{document}